\documentclass[twocolumn,aps,prl,superscriptaddress, amsmath,amssymb]{revtex4-2}
\usepackage{graphicx}
\usepackage{dcolumn}
\usepackage{bm}
\usepackage{epstopdf}
\usepackage{xcolor}
\usepackage{siunitx} 

\newcommand{\Bo}[1]{{\color{black} #1}}

\usepackage[normalem]{ulem} 

\begin{document}

\title{Hyperuniform  Active Chiral Fluids with Tunable Internal Structure}

\author{Bo Zhang}
\affiliation{Materials Science Division, Argonne National Laboratory, 9700 South Cass Avenue, Lemont, IL 60439, USA}

\author{Alexey Snezhko}
\email{snezhko@anl.gov}
\affiliation{Materials Science Division, Argonne National Laboratory, 9700 South Cass Avenue, Lemont, IL 60439, USA}

\date{\today}
\begin{abstract}

Large density fluctuations observed in active systems and hyperuniformity are two seemingly incompatible phenomena. However, the formation of hyperuniform states has been recently predicted in non-equilibrium fluids formed by chiral particles performing circular motion with the same handedness. Here we report evidence of hyperuniformity realized in a chiral active fluid comprised of pear-shaped Quincke rollers of arbitrary handedness. We show that hyperuniformity and large density fluctuations, triggered by dynamic clustering, coexist in this system at different length scales.  The system loses its hyperuniformity as the curvature of particles' motion increases transforming them into localized spinners. Our results experimentally demonstrate a novel hyperuniform active fluid and provide new insights into an interplay between chirality, activity and hyperuniformity.

\end{abstract}

\maketitle


Disordered hyperuniform materials constitute a new class of correlated systems that suppress the long-wavelength density fluctuations similar to perfect crystals or quasicrystals while still resembling the isotropic structures in disordered systems such as liquids or glasses \cite{torquato2018hyperuniform, dreyfus2015diagnosing}.
The isotropic nature and defect insensitivity of the disordered hyperuniform materials make them promising candidates for applications in photonics as materials with isotropic complete photonic band gaps \cite{florescu2009designer, man2013isotropic}.
The disordered hyperuniform states have been observed in a handful of systems, including  periodically driven colloidal suspensions~\cite{tjhung2015hyperuniform, weijs2015emergent, weijs2017mixing, wang2018hyperuniformity, wilken2020hyperuniform}, jammed packings \cite{donev2005unexpected, berthier2011suppressed, zachary2011hyperuniform, dreyfus2015diagnosing, martelli2017large, ricouvier2017optimizing, ma2020optimized}, block copolymers assemblies \cite{zito2015nanoscale, chremos2018hidden}, amorphous 2D materials \cite{zheng2020disordered, chen2021stone}, biological systems \cite{jiao2014avian, huang2021circular}, and simulation models \cite{hexner2015hyperuniformity, hexner2017noise, hexner2017enhanced, lei2019hydrodynamics, lei2019nonequilibrium, oppenheimer2022hyperuniformity}.

Active matter, comprised of active units autonomously transducing energy from the environment or delivered by external fields into a mechanical motion, exhibits  a remarkable variety of out-of-equilibrium  self-organization and coherent motion \cite{marchetti2013hydrodynamics,sanchez2012spontaneous,vicsek2012collective,snezhko2011magnetic, palacci2013living, sokolov2012physical, snezhko2016complex, soni2019odd, han2021fluctuating}  often accompanied by large density fluctuations \cite{narayan2007long}. The giant number fluctuations and spontaneous dynamic phase separation usually observed in active systems  supposedly should prevent the formation of hyperuniform states in these systems. Nevertheless,  it was recently predicted~\cite{lei2019nonequilibrium} that those seemingly incompatible properties can, in principle, coexist in certain active systems albeit at different length-scales.  It was demonstrated in simulations that in a model system of active particles preforming independent circular motions with the same handedness  a non-equilibrium hyperuniform fluid phase may form with vanishing long-wavelength density fluctuations~\cite{lei2019nonequilibrium}. While the global hyperuniformity in that system was observed at large length-scales, large local density fluctuations were confined within the length-scale comparable to the radius of the circular motion of the active particles. Recently, the existence of a disordered hyperuniform state in a system of marine algae performing circular motion was reported \cite{huang2021circular}. However, the large density fluctuations were fully suppressed in that system due to the lack of velocity alignment mechanisms.


In this letter we report a first experimental realization of a disordered hyperuniform fluid with large local density fluctuations realized in an active colloidal ensemble. We employ pear-shaped  particles energised by Quincke electro-hydrodynamic phenomenon~\cite{quincke1896ueber} to achieve spontaneous motion of particles at circular trajectories with radii controlled by the strength of the external electric field, and handedness spontaneously selected by each particle upon application of the field. We demonstrate that dynamic self-organization in such active ensembles leads to the formation of a hyperuniform fluid phase with significantly suppressed long-wavelength density fluctuations approaching those in crystal structures.  We show that the global hyperuniformity in our system often coexists with large local density fluctuations at intermediate length-scales associated with the formation of dynamic clusters, such as localized vortices or flocks. We further demonstrate that hyperuniformity disappears  in those ensembles as the curvature of particles' motion increases and a circular motion transforms into a localized spinning,  reducing the interactions between the particles and preventing  the formation of collective states.
In a scenario, when a particle motion is no longer chiral, realized by spherical Quincke rollers, the hyperuniformity is also not observed. Our findings highlight a surprising interplay between the chirality of particle motion and alignment interactions that leads to the formation of disordered hyperuniform states in active fluids.




\begin{figure*} [!htbp]
\centering
\includegraphics[width=1.0\linewidth]{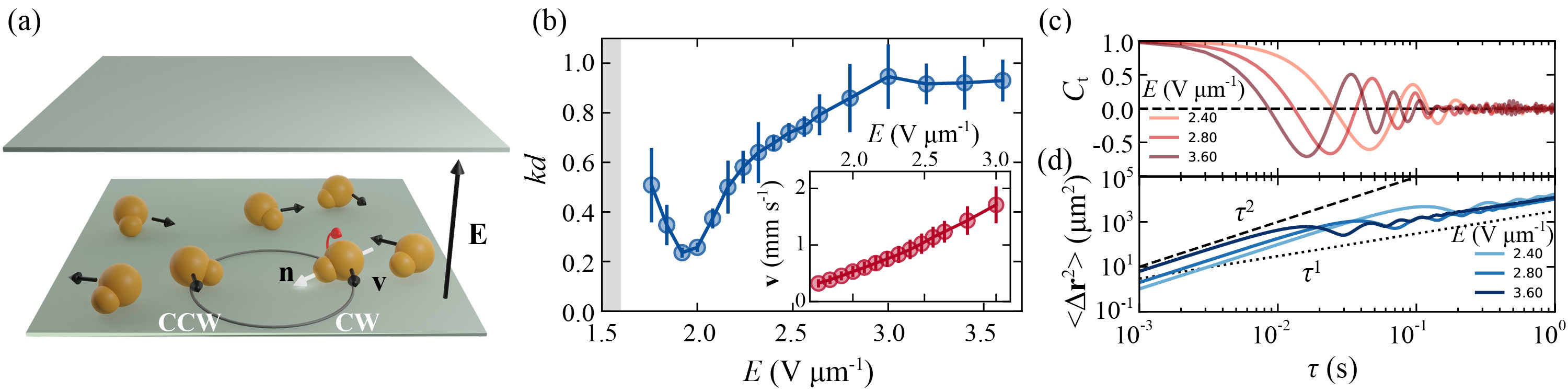}
\caption{
 	(a) A sketch of the experimental setup. The red arrow depicts the rotational motion and the black arrows show the direction of the translational motion of the pear-shaped particles energised by an external DC electric field \textbf{E}. The white arrow indicates the particle orientation $\textbf{n}$. The gray circle illustrates a trajectory of a roller that can be either in CW or CCW state.
	(b) The curvature $k$ of particles' trajectories in a dilute sample obtained at different field strengths. Particles do not perform steady rotations when $E<E_\text{c}=$ 1.6 V \SI{}{\micro\meter}$^{-1}$ (shown as gray background).
	Inset: Average particle velocity $\bf{v}$ of the rollers in a dilute sample as a function of the electric field strength.
	The area fraction $\phi =$ 0.011.
	(c) Velocity temporal correlation function $C_\text{t}$ in a dilute suspension. $\phi =$ 0.011.
	(d) Mean square displacements of circle rollers at different field strengths. The dotted and dash lines indicate $\tau^1$ and $\tau^2$ dependencies, respectively. $\phi =$ 0.116.
	}
\label{Fig1}
\end{figure*}

In our experiments, suspensions of pear-shaped polystyrene particles (PNT010UM, Magsphere Inc; long axes $d_\text{l}=$ 10.5 \SI{}{\micro\meter} and short axes $d_\text{s}=$ 9.0 \SI{}{\micro\meter}; the average particle size $d= (d_\text{l}+d_\text{s})/2$) in a 0.15 mol L$^{-1}$ AOT/hexadecane solution are sandwiched between two parallel ITO-coated glass slides and energized by a static (DC) electric field applied perpendicular to the glass slides, see Fig.~\ref{Fig1}(a). Above a certain threshold value of the field strength, $E_\text{c}$, pear-shaped particles start to spontaneously rotate due to electro-hydrodynamic Quincke rotation phenomenon \cite{tsebers1980internal, zhang2020reconfigurable} and turn into rollers exploring the bottom plate of the experimental cell.

Similar to the case of spherical Quincke rollers~\cite{bricard2013emergence,pradillo2019quincke,zhang2020oscillatory} the individual particle's speed (activity) is controlled by the  strength of the external electric field, see the insert of Fig.~\ref{Fig1}(b). However, the shape-anisotropy of the pear-shaped rollers  results in a significant change in the mechanics of the particle rolling \cite{zhang2020reconfigurable}. In contrast to spherical Quincke rollers spontaneously selecting an axis of the rotation in the plane transversal to the external electric field, a pear-shaped roller favors rotations around its long-axis [$\textbf{n}$ in Fig.~\ref{Fig1}(a)] due to a viscous drag anisotropy. Consequently, the pear-shaped rollers translate in the direction orthogonal to the long-axis. In addition, as the particle is also anisotropic along the long axis, the trajectory of the rolling motion becomes curved\cite{zhang2020reconfigurable}. The chiral state of each pear-shaped roller is spontaneously selected during system activation by the electric field, and  both types of a chiral motion, clockwise(CW) and counter-clockwise (CCW), are simultaneously realized in the system with equal probability, see Fig. S1 \cite{SI}. In general, the orientation of the long-axis of pear-shaped rollers with respect to the bottom surface (tilt) depends on the electric field strength~\cite{zhang2020reconfigurable} that provides a valuable control over the curvature of the rollers' trajectories as demonstrated in Fig.~\ref{Fig1}(b) where the curvature, $k$, of the rollers' trajectories is plotted as a function of the external field strength $E$. The behavior of the curvature is non-monotonic with the field strength due to different modes of the particle rolling discussed previously in Ref.~\cite{zhang2020reconfigurable}, yet it enables in-situ manipulation of the particles circular motion.

The change in a circular motion of the pear-shaped rollers can be quantified by changes in the velocity temporal correlation function, $C_\text{t}(\tau) = N^{-1} \sum_i { \langle\textbf{v}_i(t) \cdot \textbf{v}_i(t+\tau) \rangle}_t/{ \langle \textbf{v}^2_i(t) \rangle}_t$ of the rollers, shown in Fig.~\ref{Fig1}(c). Here, $\textbf{v}_i$ is the velocity of a roller $i$; $\tau$ is the time interval between the observations; $N$ is the total number of rollers; $\langle \; \rangle_{t}$ indicates time average. As the field strength $E$ increases, the period of oscillations of the $C_\text{t}$ decreases due to a gradual increase of the particles' velocities and curvature of their  trajectories. When in ensembles pear-shaped circle rollers develop tendril-like trajectories due to interactions with neighbouring particles with characteristic mean square displacement (MSD), $\langle \Delta \textbf{r}^2(\tau) \rangle =  \langle [ \textbf{r}(t+\tau) - \textbf{r}(t) ]^2\rangle$, curves shown in  Fig.~\ref{Fig1}(d). The oscillations in the MSDs reflect circular motion of the rollers. The behavior of the particles is initially ballistic with $\langle \Delta \textbf{r}^2(\tau) \rangle \sim \tau^2$, that eventually transitions to a diffusive regime. The transition to a diffusive regime occurs faster for rollers with higher curvatures of their trajectories corresponding to the higher strengths of the field.
\Bo{The active diffusion constant determined by the integral over the velocity autocorrelation function \cite{belkin2009magnetically}, $D = 1/2 \int_{0}^{\infty}  \textbf{v}(t_0) \cdot \textbf{v}(t_0+t) \text{d}t$, closely coincides with one obtained from the diffusive slope of the MSD curves, see Fig. S3 \cite{SI}.}
Overall, the ensemble of pear shaped rollers is a robust experimental platform to realize a 2D active system of independent circle active particles that can perform motion with random handednesses and circling phases.

\begin{figure} [!htbp]
\centering
\includegraphics[width=1.0\linewidth]{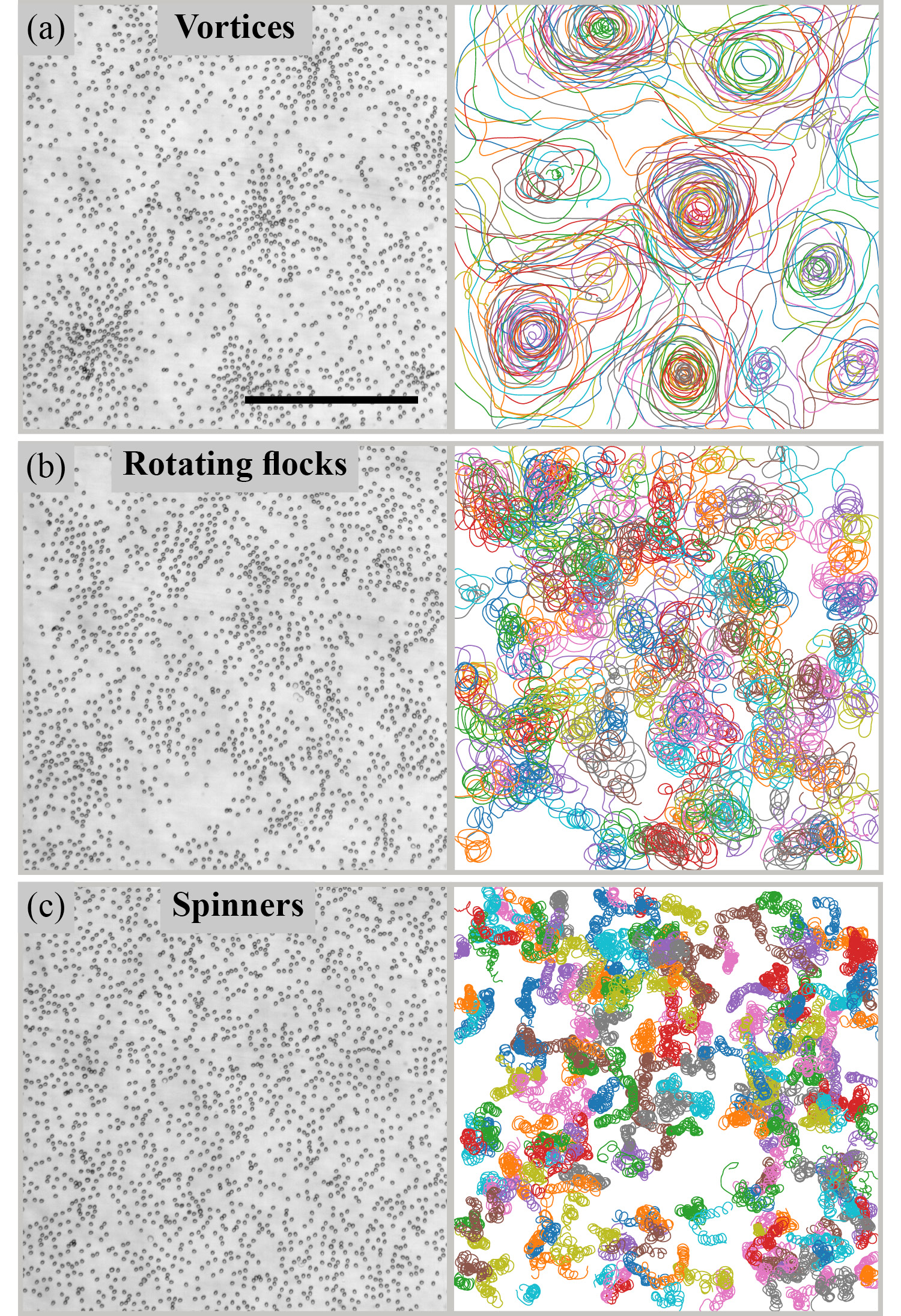}
\caption{ Snapshots of collective emergent patterns (left) and corresponding trajectories (right)  formed by circle rollers.
	(a) Spontaneous localized vortices. $E=$ 2.0 V $\SI{}{\micro\meter}^{-1}$.
 (b)Rotating flocks. $E=$ 2.4 V $\SI{}{\micro\meter}^{-1}$.
 (c) Gas of spinners. $E=$ 3.2 V $\SI{}{\micro\meter}^{-1}$.
	Only a parts of the whole experimental images and  10 \% of the corresponding trajectories are shown for clarity.
	The scale bar is 0.5 mm.  $\phi =$ 0.116.
	}
\label{Fig2}
\end{figure}

Dense ensembles of pear-shaped rollers have a strong tendency towards  spontaneous formation of dynamic patterns. \Bo{The rolling motion of the particles is crucial to maintain the desired collective patterns, in contrast to hovering behavior \cite{pradillo2019quincke} reported previously at field strengths below the Quincke rotation threshold. } Formation of collective phases in a system of circle rollers is promoted by velocity alignment interactions (facilitated by hydrodynamic and electrostatic interactions among active particles) \Bo{and it is very sensitive to the particle number density}~\cite{zhang2020reconfigurable}. A set of remarkable dynamic phases has been recently revealed in this system~\cite{zhang2020reconfigurable} with most prominent emergent patterns shown in Fig.~\ref{Fig2} accompanied by representative trajectories of the circle rollers forming the dynamic states.
At a low electric field strength ($E <$ 2.3 V $\SI{}{\micro\meter}^{-1}$), rollers form multiple vortices with well defined average sizes that can be manipulated by the electric field strength~\cite{zhang2020reconfigurable, zhang2021persistence}, rollers persistently rotate around their corresponding vortical centers, see Fig.~\ref{Fig2}(a) and Video S1,S2 in \cite{SI}. As the field strength $E$ increases  the interaction range of the individual rollers decreases (since the curvature of the trajectories increases, rollers become more localized) resulting in gradual transformation of vortices into rotating flocks \cite{zhang2020reconfigurable}. Rollers belonging to the same flock synchronize their velocity and phases with typical tendril-like trajectories shown in Fig.~\ref{Fig2}(b), see also Video S3 in \cite{SI}. Further decrease of the interaction length of the circle rollers driven by the increase of the  curvature of their trajectories results in the loss of synchronization between the circle rollers and the formation of a spinner gas phase, see Fig.~\ref{Fig2}(c) and Video S4 in \cite{SI}.


\begin{figure*} [!htbp]
\centering
\includegraphics[width=1.0\linewidth]{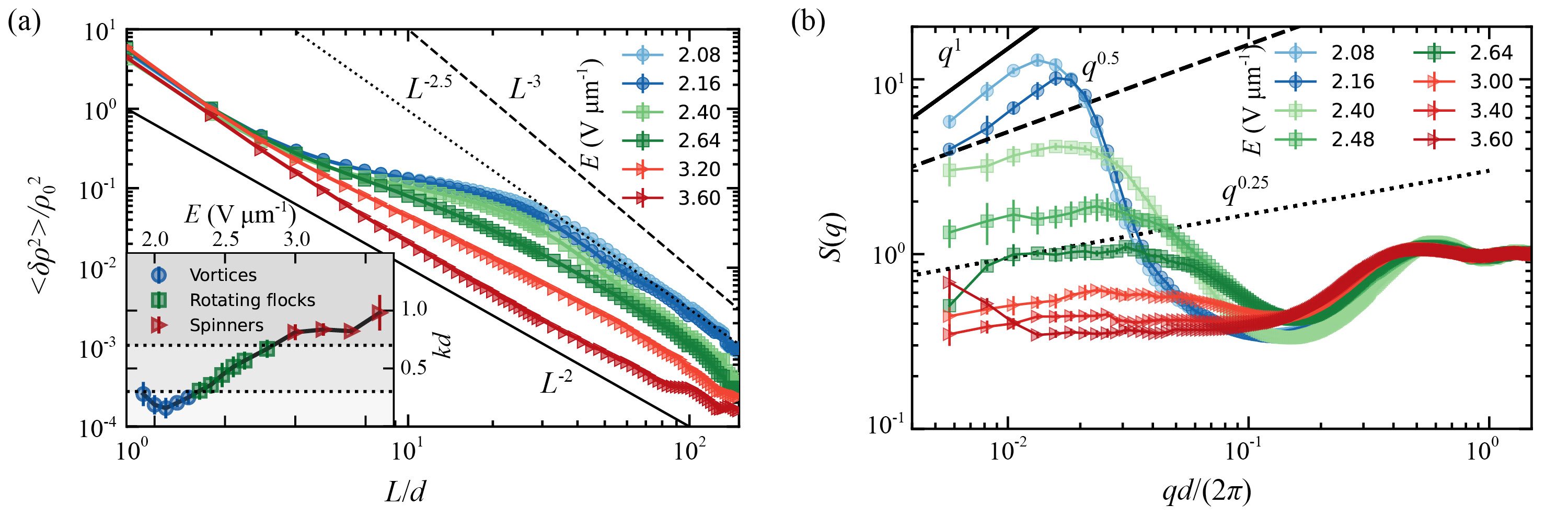}
\caption{
Dynamic hyperuniform states with local density fluctuations.
 	(a) Density fluctuations $\langle \delta \rho(L)^2 \rangle$ of pear-shaped rollers as a functions of the window size $L$. The solid, dotted and dashed lines show $L^{-2}$, $L^{-2.5}$ and $L^{-3}$ scalings, respectively.
	Inset: Curvatures of particle trajectories at different electric field strengths.
 	(b) Structure factors $S(q)$ of circle rollers at different dynamic states. Solid, dashed  and dotted lines show $q^{1}$, $q^{0.5}$ and $q^{0.25}$ scalings, respectively.
	The shapes and color of the symbols correspond to different dynamic  phases: vortices (blue circles), rotating flocks (green squares) and spinners (red triangles).
	The area fraction $\phi =$ 0.116.
}
\label{Fig3}
\end{figure*}

\begin{figure} [!htbp]
\centering
\includegraphics[width=1.0\linewidth]{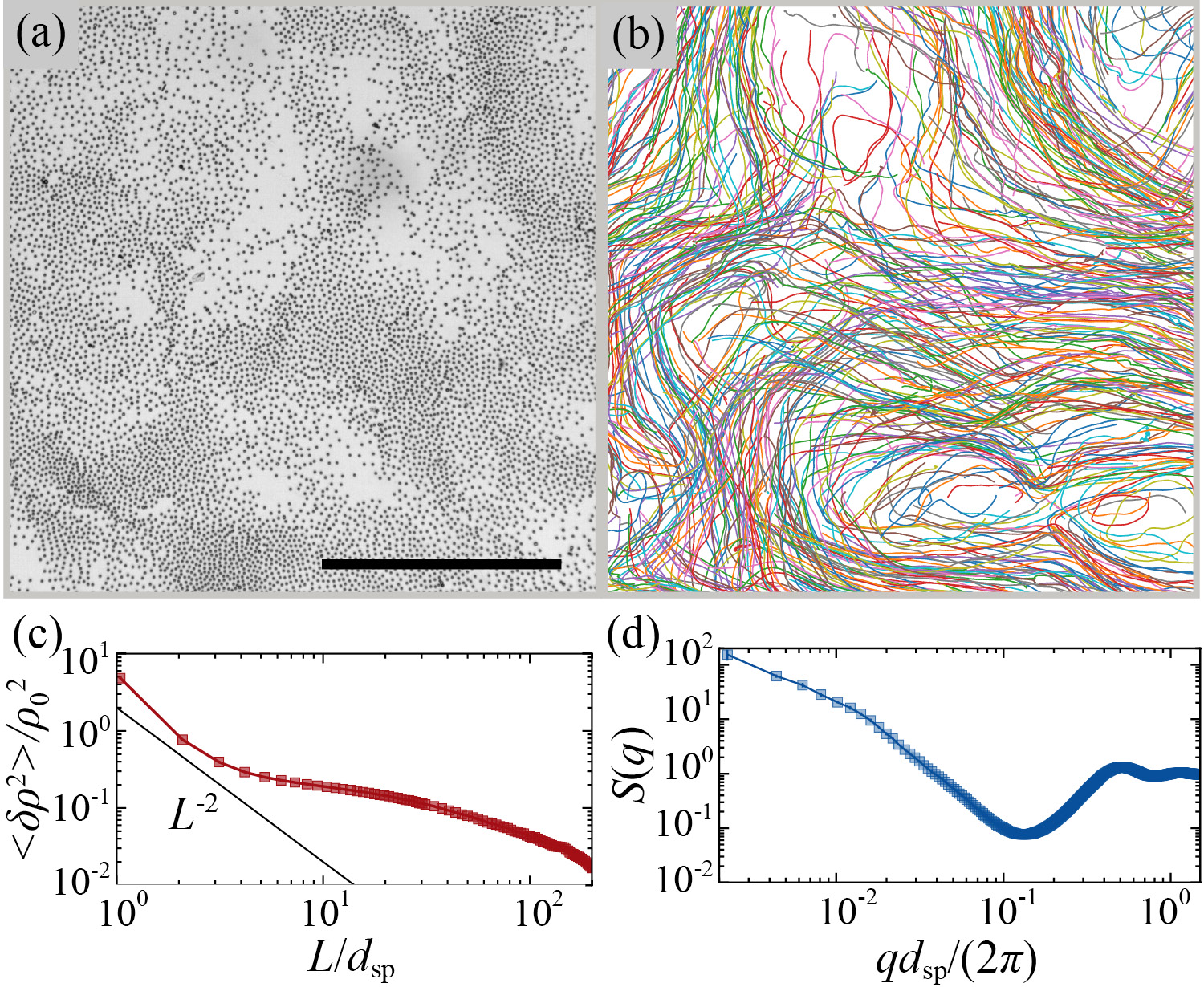}
\caption{
Flocks of spherical rollers powered by a pulsed electric field.
	(a) A snapshot of a flocking state of spherical rollers.
 (b)trajectories of rollers in flocks shown in (a). Only part of the whole experimental image and 10 \% of the  trajectories are shown for clarity. The scale bar is 0.5 mm. The electric field strengths is 2.5 V $\SI{}{\micro\meter}^{-1}$; The area fraction $\phi =$ 0.124. The frequency and duty cycle of the pulsed electric field are 100 Hz and 85 \%, respectively.
	See also Video S5 in \cite{SI}.
	(c)-(d) Local density fluctuations $\langle \delta \rho(L)^2 \rangle$ as a functions of the window size $L$ and structure factor, $S(q)$, of the rollers. The size of spherical particles $d_\text{sp} =$ 4.8 $\SI{}{\micro\meter}$.
	}
\label{Fig4}
\end{figure}

To quantify the structural properties and demonstrate the emergence of hyperuniformity in suspensions of active circle rollers, we first examine the density fluctuations, $\langle \delta \rho ^2\rangle$, defined as $\langle \delta \rho(L)^2\rangle = \langle \rho(L) ^2\rangle - \langle \rho(L) \rangle^2$, where $\rho(L)$ is the particle number density in a region with the window size $L$. In regular disordered materials such as gases, liquids or amorphous solids, the density variance scales as $\langle \delta \rho(L) ^2\rangle \sim L^{-\lambda}$ with $\lambda = D$ \cite{torquato2018hyperuniform}; $D$ here is the dimensionality of the system. When $\lambda > D$ the particles are distributed more uniformly and the structure is hyperuniform. Many active systems develop large density fluctuations~\cite{narayan2007long,dey2012spatial} characterized by enhanced density variance with $\lambda < D$.

In Fig.~\ref{Fig3}(a) the particle number density fluctuations scaling is shown for  three dominant dynamic phases in the ensemble of pear-shaped circle rollers. Collective phases, represented by localized vortices and rotating flocks, exhibit large density fluctuations (green and blue curves in Fig.~\ref{Fig3}(a)) below a certain length-scale, $L_c \sim $30d, corresponding to a typical size of the dynamic clusters in the system, either vortices or rotating flocks. The gas of uncorrelated spinners, on the other hand, does not show enhancement in density fluctuations, and its density variance  follows normal disordered fluids behavior, $\langle \delta \rho(L) ^2\rangle \sim L^{-D}, D =2$ (red curves in Fig.~\ref{Fig3}(a)).

Above $L_c$ the system exhibits clear hyperuniform scaling in the regime of vortices and rotating flocks, with $\langle \delta \rho(L) ^2\rangle \sim L^{-2.5}$, see Fig.~\ref{Fig3}(a). As the transition between dynamic phases is manipulated by the control of the curvature of the circle rollers, the hyperuniform scaling in the ensemble of pear-shaped rollers is observed for $kd < 0.7$ (see insert of Fig.~\ref{Fig3}(a)) that corresponds to a characteristic radius of curvature of the circle rollers $R > 1.4d$.

The evidence of the hyperuniformity in collective phases of pear-shaped circle rollers is also evident from analysis of the structure factor $S(q)$, defined as $S(q) = \langle 1/N | \sum_{j=1}^N \text{exp} (-i \textbf{q} \cdot \textbf{r}_j) | \rangle$. Here, $q= |\textbf{q}|$; $\textbf{r}_j$ is the position of particle $j$; $N$ is the total number of particles. Fig.~\ref{Fig3}(b) shows the structure scaling for different dynamic phases of active rear-shaped rollers. Similarly with the data on density fluctuations scalings, the structure factors of the collective phases (localized vortices and rotating flocks) exhibit distinctive hyperuniform scaling, $S(q)\sim q^{n}\rightarrow 0$ for $q\rightarrow 0$. The strength of the hyperuniformity quantified by the exponent $n$ increases with the  increase of the radius of curvature and reaches maximum for the vortex phase ($n\sim 1$), while for the case of the rotating flocks the scaling is less pronounced, $S(q)\sim q^{0.25}, q\rightarrow 0$, see Fig.~\ref{Fig3}(b).
Different values of the hyperuniform scaling exponent $n$ have been previously reported in several experimental 2D soft matter systems, such as in sheared colloidal suspensions ($n$ = 0.25) \cite{wilken2020hyperuniform}, in periodically driven emulsions ($n$ = 0.5) \cite{weijs2015emergent} and in algae suspensions ($n$ = 0.6) \cite{huang2021circular}.

The coexistence of hyperuniformity and large density fluctuations in collective phases of active pear-shaped rollers is manifested by a pronounced peak in the structure factor [green and blue curves in Fig.~\ref{Fig3}(b)] corresponding to a dynamic clustering of the circle rollers (into vortices or rotating flocks). The crossover from a large density fluctuations regime to a hyperuniform scaling takes place at length-scales corresponding to a characteristic size of the dynamic clusters reflected by the shift of the peak in $S(q)$ with the change of the curvature of the circle rollers motion. This findings are important as they experimentally confirm that hyperuniformity and large density fluctuations in active chiral systems can in principle coexist at different length-scales in accordance with the predictions in Ref.~\cite{lei2019nonequilibrium} (although obtained for the system of circle active particles with the same handedness). Our results demonstrate that identical handedness of the circle active particles is not required for the coexistence of those two phenomena and systems with heterogeneous handedness of the active particles share similar phenomenology.

\Bo{The formation of hyperuniform states is sensitive to the particle number density. The degree of hyperuniformity and large density fluctuations decrease with the  area fraction of the chiral rollers (see Fig.\~S4 in \cite{SI}), indicating the significance of collective interactions for the onset of hyperuniformity in this system.}

The importance of the circular motion of the rollers for the emergence of hyperuniformity  is further demonstrated by the absence of hyperuniform structures in flocks formed by spherical rollers driven by a  pulsed electric field~\cite{karani2019tuning,zhang2022guiding}. As Fig.~\ref{Fig4} shows, while the large density fluctuations are still clearly observed in the density variance $\langle \delta \rho (L)^2 \rangle$ (see Fig.~\ref{Fig4}(c)) and in the structure factor $S(q)$ of the spherical rollers (Fig.~\ref{Fig4}(d)), the hyperuniform state no longer exists.


In summary, active ensembles of pear-shaped Quincke rollers performing circular motion with arbitrary handedness realize a
disordered hyperuniform active fluid in a certain range of curvatures of the particles' motion. Our results experimentally demonstrate that the global hyperuniformity in active chiral systems can coexist with large density fluctuations at intermediate length-scales associated with the dynamic clustering. The hyperuniformity disappears  in those ensembles as the curvature of particles' motion increases and a circular motion transforms into a localized spinning,  reducing the interactions between the particles and preventing  the formation of collective states.  The absence of the hyperuniformity in ensembles of flocking spherical rollers emphasizes the importance of a chiral circular motion of active particles  for a formation of self-organized anomalously homogeneous active structures. This surprising interplay between chirality, activity and hyperuniformity may be used as a tool to assemble materials into disordered hyperuniform structures.



The research was supported by the U.S. Department of Energy, Office of Science, Basic Energy Sciences, Materials Sciences and Engineering Division.

\bibliography{Refs}
\bibliographystyle{apsrev4-2}  


\end{document}



\date{}
\maketitle

\clearpage


\begin{figure*}[!htbp]
\centering
\includegraphics[width=1.0\textwidth]{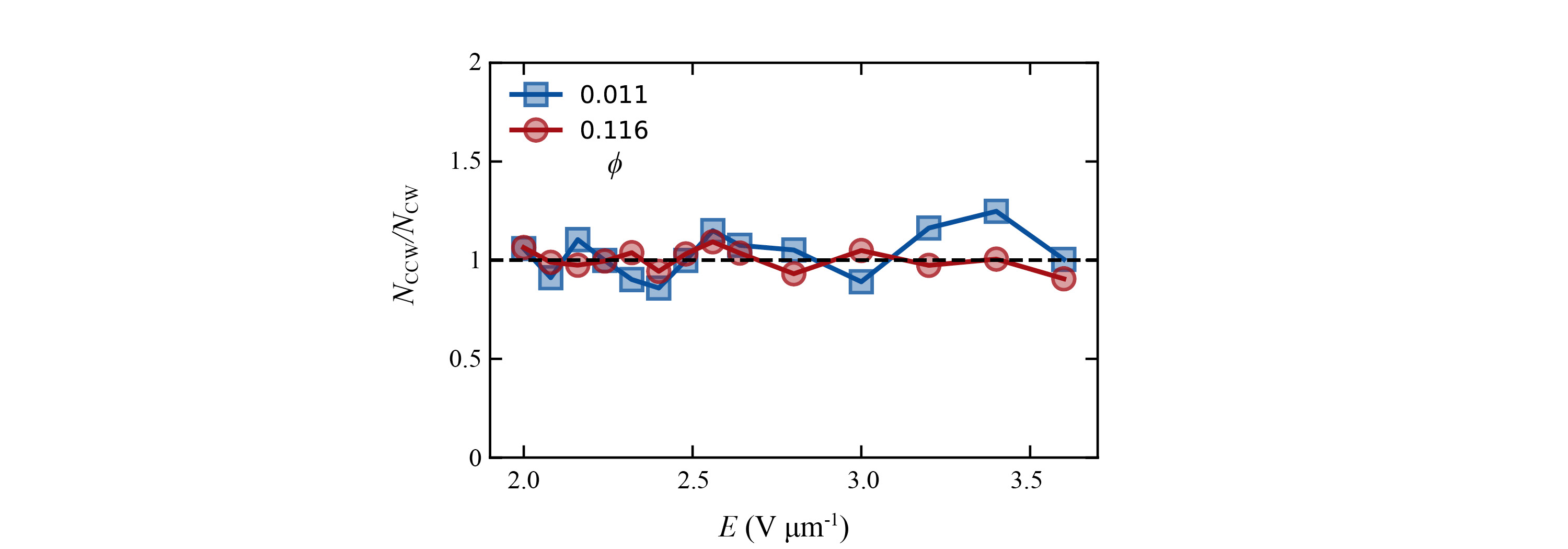}
\caption{
	The ratio between the numbers of counter clockwise $N_\text{CCW}$ and clockwise $N_\text{CW}$  rollers. The ratio is about 1 to 1, indicating that both chiralities of rollers are equally represented in the system regardless of the field strengths or the area fraction. Small fluctuations are observed due to the limited field of view.
}
\label{FigS1}
\end{figure*}

\begin{figure*}[!htbp]
\centering
\includegraphics[width=1.0\textwidth]{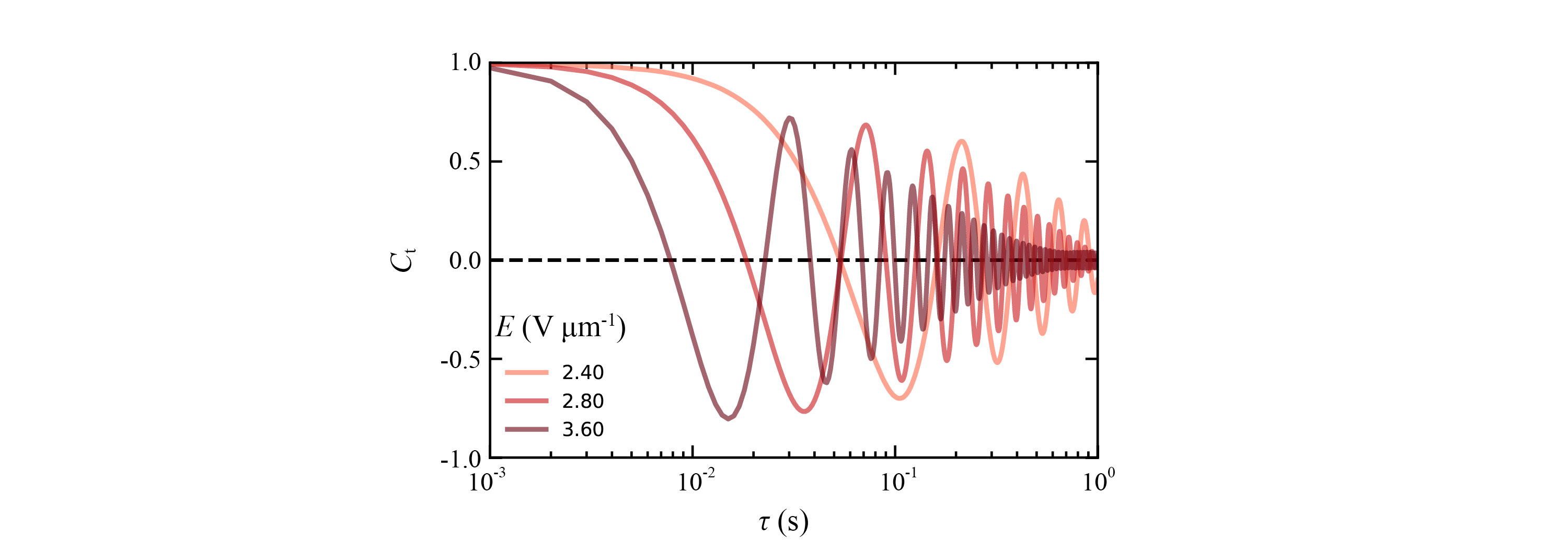}
\caption{
	Velocity temporal correlation function $C_\text{t}$ at different field strengths. The area fraction $\phi =$ 0.116. The oscillations decay slower than those in dilute samples due to collective effects.
}
\label{FigS2}
\end{figure*}

\begin{figure*}[!htbp]
\centering
\includegraphics[width=1.0\textwidth]{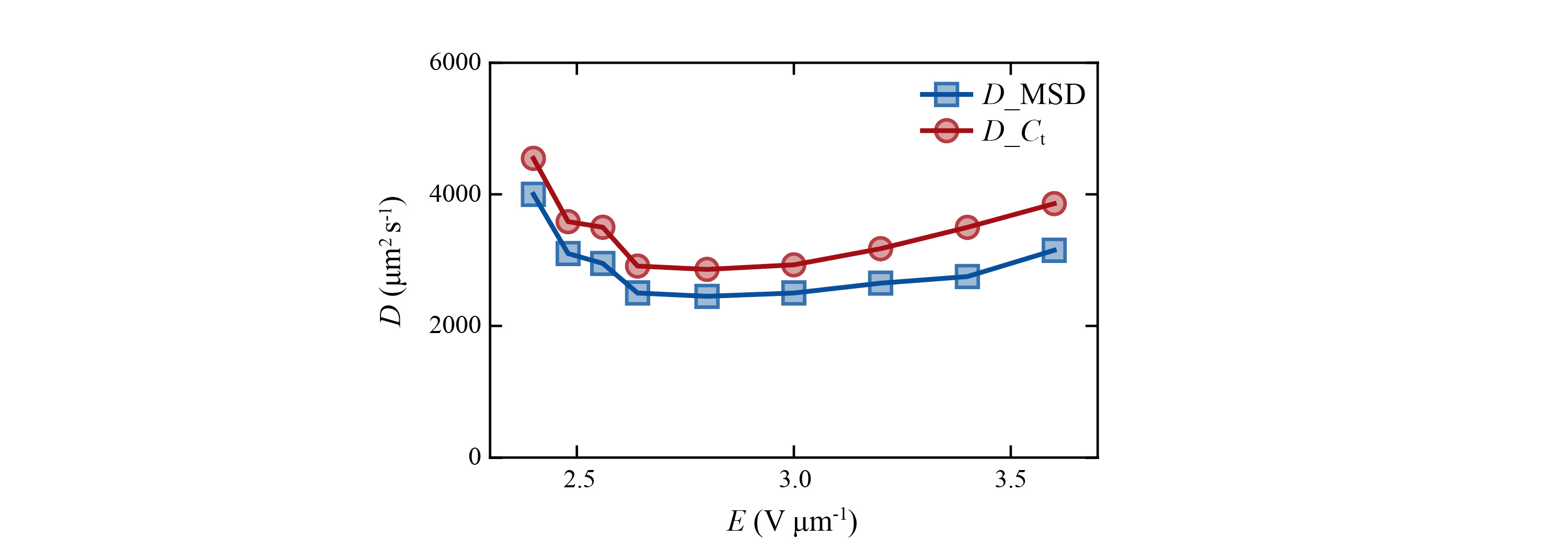}
\caption{
	Diffusion coefficient constants, $D$, measured from diffusive regions of MSD curves (blue squares, $D= \langle \Delta\bf{r}^2\rangle / (4\tau)$) and integrals of velocity temporal correlation functions (red circles, $D = 1/2 \int_{0}^{\infty}  \textbf{v}(t_0) \cdot \textbf{v}(t_0+t) \text{d}t$). Both procedures provide very close results.
}
\label{FigS3}
\end{figure*}

\begin{figure*}[!htbp]
\centering
\includegraphics[width=1.0\textwidth]{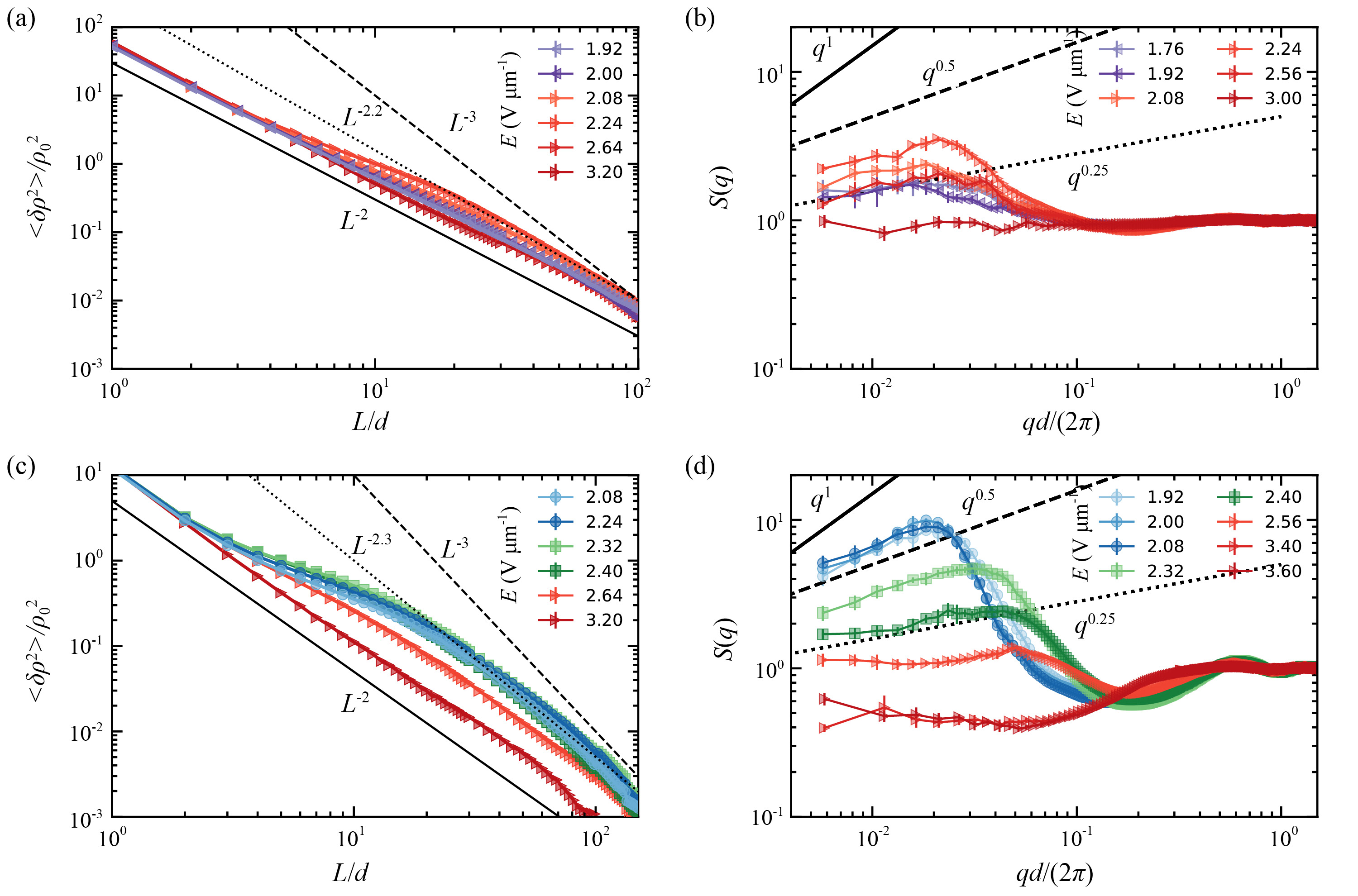}
\caption{
	Local density fluctuations $\langle \delta \rho(L)^2 \rangle$ as a functions of the window size $L$ and structure factors, $S(q)$, of the rollers. The area fractions are 0.011 and 0.049 in (a-b) and (c-d), respectively.  The shapes and color of the symbols correspond to different dynamic  phases: gas (purple left triangles), vortices (blue circles), rotating flocks (green squares) and spinners (red right triangles).
}
\label{FigS4}
\end{figure*}

\clearpage


\clearpage

\section*{Videos}

\subsection*{Video S1}
Multiple self-assembled localized vortices.
The electric field strength $E =$ 2.0 V $\SI{}{\micro\meter}$$^{-1}$.  The frame size is 1.25 mm by 1.25 mm. The movie is 0.1X of the real time.

\subsection*{Video S2}
Multiple vortices in a large field of view.
The electric field strength $E =$ 2.16 V $\SI{}{\micro\meter}$$^{-1}$.  The frame size is 5.12 mm by 5.12 mm. The movie is 0.1X of the real time.

\subsection*{Video S3}
Rotating flocks.
The electric field strength $E =$ 2.4 V $\SI{}{\micro\meter}$$^{-1}$.  The frame size is 1.25 mm by 1.25 mm. The movie is 0.1X of the real time.

\subsection*{Video S4}
Spinners.
The electric field strength $E =$ 3.2 V $\SI{}{\micro\meter}$$^{-1}$.  The frame size is 1.25 mm by 1.25 mm. The movie is 0.1X of the real time.

\subsection*{Video S5}
Flocks of spherical rollers energized by a pulsed electric field.
The electric field strength $E =$ 2.5 V $\SI{}{\micro\meter}$$^{-1}$. The frequency is 100 Hz and the duty cycle is 85 \%.  The frame size is 1.25 mm by 1.25 mm.  The movie is 0.2X of the real time.